\documentclass[lnbip]{svmultln}
\usepackage[T1]{fontenc}
\usepackage{graphicx}

\usepackage{hyperref}

\hypersetup{
  colorlinks=true,
  linkcolor=blue,
  citecolor=blue,
  urlcolor=blue
}

\usepackage{xcolor}

\usepackage{xspace}
\newcommand{\teammate}{SE agent\xspace}
\newcommand{\approach}{\textsc{pm$_4$aa}\xspace}
\usepackage[most]{tcolorbox}
\usepackage{subcaption} 
\usepackage{cleveref}
\usepackage{microtype}

\usepackage{pifont}

\tcbset{
  pmbox/.style={
    colback=gray!30, colframe=gray!90,
    left=3pt, right=3pt, top=3pt, bottom=3pt,
    before skip=4pt, after skip=4pt
  }
}
  
\begin{document}

\title{Using Process Mining to Generate AI Agents from Software Engineering Process Records}
\titlerunning{Process Mining to Generate AI Agents}


\author{Saimir Bala\inst{1} \and Fabiana Fournier\inst{2,3} \and Lior Limonad\inst{2,3} \and Andreas Metzger\inst{4}}
\institute{
  Hasso Plattner Institute (HPI), University of Potsdam, Germany \\
  \texttt{saimir.bala@hpi.de}
\and
  IBM Software Innovation Lab (SIL), Haifa \\
  \texttt{fabiana@il.ibm.com, liorli@il.ibm.com}
\and
  Dept. of Information Systems,
  Fac. of Comp. \& Inform. Science,
  University of Haifa
\and
  paluno Institute, University of Duisburg Essen, Essen Germany \\
  \texttt{andreas.metzger@paluno.uni-due.de}
}

\maketitle
\begin{abstract}
Integrating AI agents into Software Engineering (SE) raises an important challenge: how can we specify and realize AI agents that work effectively alongside humans in hybrid SE teams?
Determining the right granularity and separation of concerns for such agents is non-trivial.
Coarse-grained agents may introduce unmanageable complexity, whereas micro-agents may create severe coordination overhead.
Moreover, existing multi-agent SE frameworks typically rely on predefined role structures and do not account for project-specific characteristics or process adaptations.
We address this by combining object-centric, imperative, and declarative process mining.
Using event logs extracted from software repositories, our approach discovers project-specific agent roles using a predefined SE role vocabulary grounded in repository behavior and generates matching agent specifications and implementations.
As proof-of-concept, we applied our approach to a well-established open-source project.  
We performed functional tests and 
an exploratory user study to determine how well the generated AI agent specifications are aligned with human expectations.
\end{abstract}


\section{Introduction}
\label{sec:intro}

Process mining turns event data into insights about how work is performed, who performs it, and under which constraints~\cite{van2016data}.
Software engineering (SE) processes are no exception: the activities of developers, reviewers, and testers leave rich event traces in version control and issue management systems, forming a largely untapped source of process knowledge.
Today, SE teams are undergoing a fundamental transformation as AI agents (autonomous systems capable of decomposing goals, executing tasks, and collaborating with humans) are increasingly embedded into development workflows~\cite{RoychoudhuryPPR26,TOSEM2026,OtoumE26}.
The SE community has already acknowledged the rise of \emph{AI teammates}, and the evolution towards \emph{hybrid human-AI SE teams} raises an important new engineering challenge: \textit{How to specify and realize AI agents that can effectively work alongside humans?}

Addressing this question has several challenges.
Assigning a classical SE role (such as \emph{team member} in agile processes) to a single agent yields a monolithic and complex teammate that is difficult to build, test, maintain, and delegate tasks to.
Conversely, defining highly fine-grained, single-task agents increases coordination overhead among agents and humans. 
Also, not all SE tasks may be relevant in every project, and those that are may require concrete adaptations. Formal verification may be mandatory in safety-critical systems yet irrelevant in a simple utility tool, while complex business logic may demand property-based tests beyond standard unit testing.
Finally, software engineers expect agents to adhere strictly to established standards and project-specific processes~\cite{dongtowards}.
Current frameworks rely on rigid, predefined canonical roles that do not adapt a project's actual workflows~\cite{HeTL25,RoychoudhuryPPR26,GoyalCT24,Hemmer2025}.

We address these challenges with \approach, a generative pipeline that mines project-specific SE agent specifications directly from repository event data.
\approach treats SE processes as business processes and the activity history recorded in version control and issue management systems as event logs.
It applies object-centric process mining~\cite{Seidel2026} to capture the multi-object structure of SE activities. This allows the discovery of the task scope and object interactions of each role, and declarative process mining~\cite{DBLP:books/sp/22/CiccioM22} to elicit behavioral constraints that serve as guardrails for agent execution.
Together, these two complementary views provide both the procedural blueprint and the normative boundaries needed to specify a well-scoped agent.
\approach then leverages LLMs for process model analysis~\cite{KubrakBMND24} to synthesize executable agent specifications.
As a proof of concept, we implement \approach using the LangGraph framework and evaluate it through a case study on the open-source \textit{Commitizen} project, which yields an event log of more than 21,000 events spanning eight years of development history, functional testing of the generated agents, and a user study involving ten human participants, who assessed how well process-mined, project-specific agents align with human expectations in hybrid SE teams.

Below, Sec.~\ref{sec:background-and-related} introduces relevant background and related work.
Sec.~\ref{sec:approach} explains \approach.
Sec.~\ref{sec:evaluation} presents the case study and reports our \approach evaluation.
Sec.~\ref{sec:conclusion} concludes the paper and outlines future work.

\section{Background and Related Work}
\label{sec:background-and-related}

\subsection{Process Mining}
Process mining is a discipline that aims to extract insights about processes from event data traces~\cite{van2016data}. Object-Centric Process Management~\cite{Seidel2026} captures business processes around interacting objects such as orders, invoices, and shipments, rather than a single predefined workflow. Object-Centric Process Mining (OCPM) extends traditional process mining by explicitly considering processes that involve multiple interacting objects~\cite{WilOCPM2019}.
Taking object-centric event logs as input, OCPM helps capture synchronization, interaction, and dependency relations among objects and providing a more accurate
representation of highly interconnected processes.
A central artifact produced by OCPM is the Object-Centric Directly-Follows Graph (OC-DFG), which extends the classical Directly-Follows Graph (DFG) (a graphical representation of activities that directly succeed one another in the observed process).

Declarative process mining elicits process behavior through constraints that govern allowable behavior, rather than prescribing a fixed sequence of activities. Constraints express temporal or logical relations such as precedence, response, or mutual exclusion~\cite{DBLP:books/sp/22/CiccioM22,Wil2006,Maggi2011,Maggi2012}. Declarative approaches are well-suited to flexible, knowledge-intensive processes where strict control-flow specifications are difficult to define or maintain. Languages such as DECLARE have become widely adopted due to their formal semantics and expressive power, and have been applied in healthcare, customer service, and human-centric workflows.

\subsection{Event Logs from Software Repositories}
Software development is a process: developers open issues, write code, commit changes,
review pull requests, run tests, and release software.
Like a business process, it generates event data as a natural by-product of execution.
These events are captured in version control and issue management systems such as Git,
GitHub, and Jira. 

Repositories provide a rich set of data.
A single commit records the actor, a timestamp, a descriptive message, and references
to modified files, related issues, and pull/merge requests.
An issue has its own lifecycle: opened, labeled, assigned, commented upon, and
 resolved.
A pull/merge request links one or more commits to one or more issues and involves multiple
actors across review, approval, and merge events.
A single SE action, such as fixing a bug, thus spans multiple objects
simultaneously, which is precisely what OCPM is designed to capture~\cite{Seidel2026}.
\Cref{tab:ocel} shows a simplified extract of repository event data structured as an OCEL.

\begin{table}[ht]
\caption{Excerpt of a software repository structured as an Object-Centric Event Log.
A dash indicates that the event is not associated with that object type.}
\label{tab:ocel}
\centering
\small
\begin{tabular}{l l l c c c p{3.2cm}}
\hline
\textbf{Timestamp}    & \textbf{Actor} & \textbf{Activity}   &
\textbf{Commit} & \textbf{Issue} & \textbf{PR} & \textbf{Message} \\
\hline
2024-03-01 09:12 & alice & open\_issue   & {--} & \#42 & {--}  & feat: add emoji support \\
2024-03-02 14:05 & bob   & push\_commit  & a3f8c1 & \#42 & {--}  & feat(scope): add emoji support \\
2024-03-02 14:07 & bob   & open\_pr      & a3f8c1 & \#42 & \#17  & fix: handle edge case in parser \\
2024-03-03 10:30 & alice & review\_pr    & {--} & \#42 & \#17  & {--} \\
2024-03-03 11:45 & alice & merge\_pr     & a3f8c1 & \#42 & \#17  & {--} \\
2024-03-03 11:46 & bot   & close\_issue  & {--} & \#42 & \#17  & {--} \\
\hline
\end{tabular}
\end{table}

Each row represents an event tied to one or more objects of different types.
The commit, issue, and PR columns capture the multiple concurrent object lifecycles
involved in each event.
The message column carries the human-readable description attached to the change.
The table shows, for instance, how issue~\#42 spans six events across three actors:
alice opens the issue, bob commits a fix and opens a pull request, alice reviews and
merges it, and a bot closes the issue.


\subsection{Related Work on AI Agents}
\label{sec:sota}
AI agents possess their own thread of control and make autonomous decisions about which
actions to perform and when~\cite{Wooldridge2009,Li2025}. 
They decompose high-level goals into tasks and execute them in unstructured environments.
Agents can form multi-agent systems in which they collaborate with humans and other agents to carry out tasks~\cite{Wooldridge2009,Sapkota25,CALVANESE2026}. 
Their tasks are realized through AI algorithms and models, including generative AI such
as LLMs~\cite{DBLP:conf/ijcai/GuoCWCPCW024}. 
In SE, examples include autonomous coding agents that initiate, review, and evolve code
at scale~\cite{Li2025}. 

Recent literature reviews document a rapid paradigm shift from single-agent
coding assistants to complex multi-agent SE ecosystems~\cite{OtoumE26,HeTL25}.
They conclude that a single, monolithic AI agent is insufficient due to the multifaceted complexity of real-world software projects.
Consequently, state-of-the-art frameworks (e.g., ChatDev and MetaGPT) divide the development lifecycle into specialized roles, such as product managers, programmers, and testers, to improve task execution, collaborative reasoning, and fault tolerance.
However, these frameworks rely on fixed architectures
based on canonical roles (e.g., Programmer, Reviewer, 
Tester)~\cite{HeTL25}.
Such rigid role structures may cause severe inter-agent
coordination bottlenecks as projects scale~\cite{OtoumE26}.
Both~\cite{OtoumE26} and~\cite{HeTL25} call for adapting agent roles
to evolving project demands, yet operational, data-driven techniques remain lacking.
\approach directly addresses this challenge by extracting, scoping, and generating project-specific SE agents from project event logs.

The BPM field has developed organizational mining and Agent System Mining (ASM) to derive role structures and behavioral models from event logs~\cite{shen2026mining}. 
ASM tools like \textit{Agent Miner} extract complete,
multi-agent behavioral and interaction models directly from event
logs.
Still, these approaches are retrospective and simulative: they mine structures to generate synthetic traces or build digital twins of existing processes.

While \approach shares the foundational premise of Shen et al.~\cite{shen2026mining} that historical event logs contain vital, role-specific behavioral blueprints, it takes a more \emph{prescriptive} and \emph{operational} stance. Specifically, we argue that event logs contain role-specific behavior that can be mined into (OC-)DFGs and declarative constraints to engineer and deploy AI agents within an executable LangGraph architecture, enabling them to work alongside humans rather than merely simulate human actors.

\section{The PM4AA Approach}
\label{sec:approach}

\begin{figure}[h]
    \centering
    \includegraphics[width=.95\linewidth]{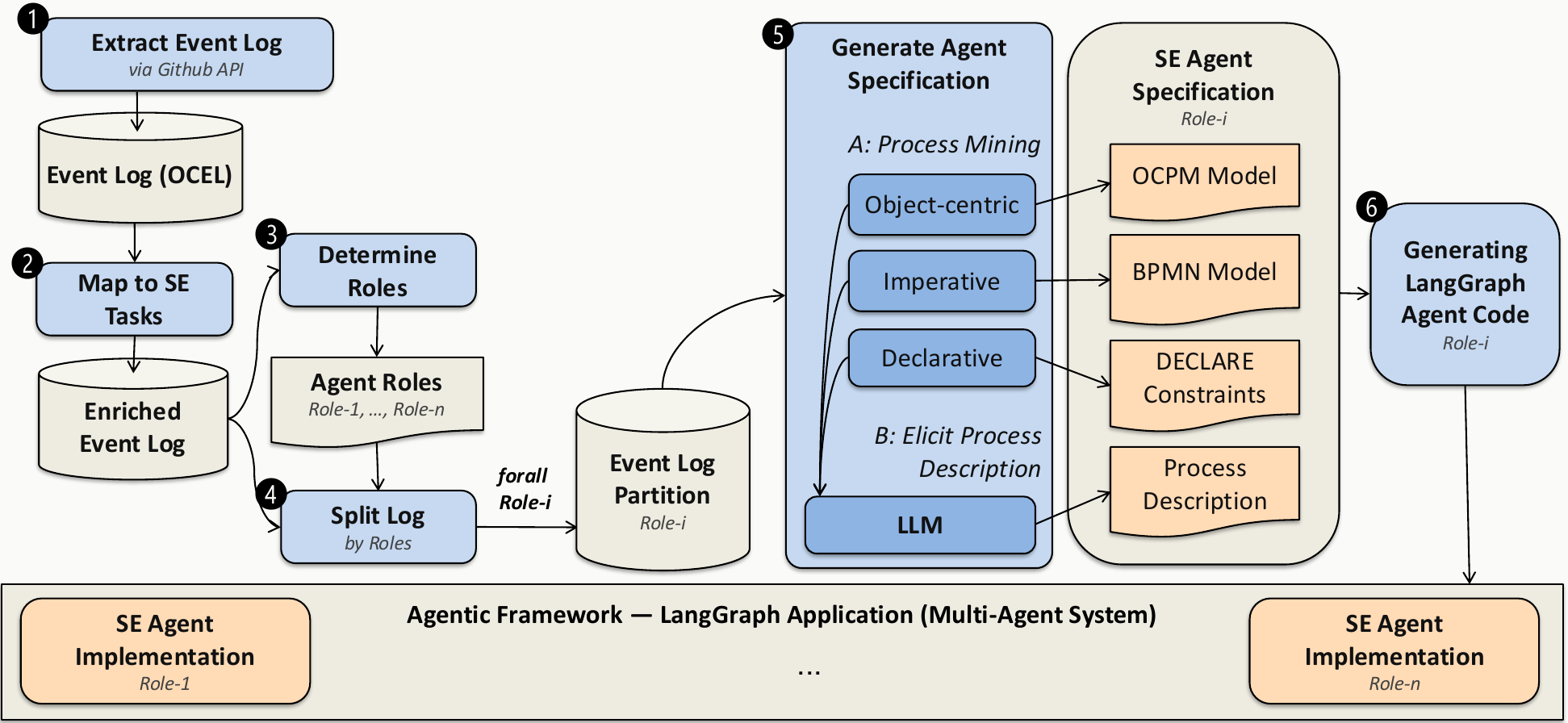}
    \caption{The \approach pipeline.}
    \label{fig:approach}
\end{figure}

\approach (Process Mining for Agentic AI) is a generic, multi-step pipeline that transforms 
SE repository data into operational, role-specific AI agents.
Fig.~\ref{fig:approach} provides an overview of the steps of the \approach pipeline, which we elaborate below.
A repository containing the full pipeline implementation and the accompanying application is available at: 
\url{https://github.com/liorlimonad/pmaa}.

\approach assumes the availability of sufficient historical repository data. It is therefore primarily applicable to existing projects with observable development history, rather than greenfield projects with no prior traces. For new projects, \approach-derived agents may be transferred only as templates and would require later adaptation once project-specific traces become available. 

\smallskip
\noindent\emph{Step \ding{202} Event Log Extraction:}
The first step is to extract a raw Object-Centric Event Log (OCEL) from a GitHub repository
using \textit{PyStack't}~\cite{Bosmans2024}, a Python library designed  to
produce OCEL~2.0 logs via the GitHub REST API.
PyStack't maps GitHub entities to OCEL object types (i.e., user account, issues and commits),  and repository interactions to
timestamped events, yielding a log that natively captures the multi-object nature of
software development activity. To derive roles, we first identify the tasks recorded in the event log.  

\smallskip
\noindent\emph{Step \ding{203} Mapping to SE Tasks:}
The raw OCEL does not represent SE \emph{tasks} as first-class objects. 
PyStack't also does not parse pull requests (often associated with tasks). Therefore, we need to derive the tasks from the commit messages.

A key enabler for this step is the commit message.
Many open-source projects enforce \textit{Conventional Commits}~\cite{ConventionalCommits},
a lightweight standard that structures commit messages as \texttt{type(scope): description}.
The type prefix classifies the intent of each change: \texttt{feat} signals a new
feature, \texttt{fix} a bug repair, \texttt{docs} a documentation update, and
\texttt{test} a testing contribution.
This allows assigning each event a corresponding type of SE task without manual
annotation~\cite{RoychoudhuryPPR26}.
Many projects (e.g., Angular, Vue.js, ESLint, and Commitizen) follow this standard.

Accordingly, we enriched the event log by parsing each commit message against the Conventional Commits
pattern \textit{type(scope): description}~\cite{ConventionalCommits} using a
regular-expression classifier.
Each successfully parsed commit is annotated with a \texttt{task\_type}
(e.g., \texttt{feat}, \texttt{fix}, \texttt{docs}) and a \texttt{task\_semantic\_class}
drawn from a fixed ten-class classification scheme derived from the Conventional Commits specification  (e.g., \texttt{feature\_development},
\texttt{defect\_resolution}, \texttt{quality\_assurance}).
A corresponding \texttt{task} object is created for each matching commit and linked
to its commit and to all events that reference that commit.

\smallskip
\noindent\emph{Step \ding{204} Determine Roles:}
Given the tasks, the goal of this step is to derive a set of process roles from the contributor behavior recorded in the OCEL (cf. organizational mining~\cite{van2016data}). All persons who have contributed to the repository (e.g., by committing a change) are available as user objects (collected in Step \ding{202}). For every user object, a \emph{resource profile} is computed by aggregating three behavioral dimensions: (i)~the distribution of activities the user participates in, (ii)~the number of commits attributed to the user, and (iii)~the distribution of task semantic classes (e.g., feature development, defect resolution, documentation) obtained from commit-level attributes in the log.

A \emph{priority-ordered rule-based} classifier then maps each profile to exactly one \emph{fixed} and \emph{prescribed} role. This hard partition is a deliberate simplification: a single dominant role per contributor keeps the sub-logs disjoint and the agent scopes unambiguous, at the cost of flattening contributors who act in multiple capacities. Organizational mining alternatively permits overlapping or weighted role assignments~\cite{van2016data}. The rules discriminate along two axes: \emph{platform identity} and \emph{behavioural dominance}. Automated accounts are identified first through their platform metadata. Contributors with no commit activity are classified as issue reporters, capturing actors whose participation is limited to issue creation and discussion. Among committing contributors, those exceeding a configurable commit-count threshold and exhibiting diversity across at least three task classes are classified as maintainers. The remaining contributors are assigned to specialist roles (such as quality engineer, DevOps engineer, technical writer, or feature developer) based on the task-class group that accounts for the plurality of their work, with a generic \emph{contributor} role as fallback. The assigned role is persisted as a new attribute on the corresponding user object in the enriched OCEL log, enabling the downstream partitioning in the next step.

\smallskip
\noindent\emph{Step \ding{205} Split Log:}
We then isolate each role's behavior.
Using the role-annotated OCEL, a per-role sub-log is constructed for each discovered role. 
For a given role, each event in which at least one user object carrying that role participates is selected. All objects referenced by those events (including non-user objects such as issues, commits, and tasks) are retained, thereby preserving the object-centric structure of the original log. The result is one OCEL sub-log per role, capturing the multi-typed object context of the activities performed by that role.


\smallskip
\noindent\emph{Step \ding{206}\textbf{-A}: Process Mining.} 
    The per-role event log partitions serve as input for three complementary process mining steps:
    
 \textit{Object-centric:}  An OC-DFG is  discovered using the \texttt{pm4py} library. 
        To filter infrequent behavior while accommodating sub-logs of varying size, a dynamic frequency threshold is applied: both the activity and edge thresholds are set to approximately 2\% of the sub-log's event count, with a minimum of one. 
        The resulting OC-DFG captures each role's typical activity sequences, annotated with the object types co-involved in each transition.
        It serves two purposes. First, it delineates the \emph{task scope} of the agent: the activities retained after thresholding determine which operations the agent is responsible for. Second, the object types in the graph define the \emph{entities the agent interacts with} (e.g., issues, commits, pull requests), grounding the agent's interface in empirical process data rather than manual specification.
    
\textit{Imperative:} Complementing the OC-DFG, a BPMN model is derived via Inductive Miner.
        The log is first flattened onto a designated primary object type (selected as the case notion), converted to a traditional event log, and the resulting process tree is then transformed into a BPMN diagram.
        Together, with the OC-DFG this provides a per-role view of \emph{how} work is carried out.

\textit{Declarative:} 
    Declarative process mining is applied to the role-specific logs to elicit normative behavioral constraints, complementing the imperative models produced previously.
    Using \texttt{DeclareMiner}~\cite{DonadelloRMS22}, each log was first flattened onto the \texttt{issue} object type to yield a case-centric representation, and constraint discovery was executed with a minimum support threshold of 0.5 and a maximum constraint cardinality of two.
    Roles with fewer than two distinct activities after flattening were excluded from discovery, as meaningful constraint inference requires activity diversity. For the remaining roles, \texttt{DeclareMiner} was executed, and the discovered constraint sets were serialised as JSON artifacts for downstream consumption.

\smallskip
\noindent\emph{Step \ding{206}\textbf{-B}: Elicit Process Description.}
    For each role, a process profile with activity frequencies, object-type distributions, directly-follows patterns, and event-object interaction counts is provided to GPT-4-mini using the following prompt.





\begin{tcolorbox}[colback=gray!30, colframe=gray!90, left=3pt, right=3pt, top=3pt, bottom=3pt] \scriptsize {\sffamily \textbf{Prompt to LLM:} Analyze the process profile for \texttt{\{role\_name\}} and generate a process description covering: (i) process overview and responsibilities, (ii) main activities grouped by function, (iii) object interactions and semantics, (iv) flow patterns and common sequences, (v) key behavioral characteristics, and (vi) business interpretation of the role's contribution to the software lifecycle. Base the analysis on an extracted process profile including total events/objects, unique activities/object types, activity frequencies, object type distributions, top directly-follows flows, and top event-object interactions. Write in professional language for business stakeholders, emphasizing object-centric behavior. } \end{tcolorbox}

The markdown output provides a structured business-oriented description of the role-specific object-centric process, including behavioral, interactional, and statistical perspectives for each of the roles.

\smallskip
\noindent\emph{Step \ding{207} Generate LangGraph Agent Code:}
With the specification in place, we move to the implementation of the agents. We make use of LangGraph
as the target
Agentic AI technology to instantiate the roles generated in the previous steps.
The LangGraph application is realized through a prompt-driven synthesis process
in which IBM BOB\footnote{\scriptsize \url{https://bob.ibm.com/}, accessed 2026-06-05} interpreted the user request and
turned the mined SE Agent specifications into an executable application.
IBM BOB employs a mixture of frontier and specialized language models
for interpretation, drawing on Anthropic Claude, Mistral open-source models, IBM Granite,
and additional specialized fine-tuned models.
The BOB prompt is as follows:

\begin{tcolorbox}[colback=gray!30, colframe=gray!90, left=3pt, right=3pt, top=3pt, bottom=3pt]
\scriptsize
{\sffamily
\textbf{$\langle$Context$\rangle$} This workspace is applied to an event log harvested from a GitHub
repository, which was then partitioned into 8 separate sub-logs
corresponding to 8 SE roles invoked in the handling of the GitHub issues. 
We derived the
Object-Centric Directly-Follows Graphs (OCDFG), BPMN graphs, DECLARE constraints and a process description markdown, for each of the logs. 

\textbf{$\langle$Task$\rangle$} Project this onto an agentic application that will facilitate and automate, where possible, online issue handling on top of the GitHub repository. 
Inspect all the attached artifacts and propose
a langgraph-based agentic application architecture matching these artifacts.}
\end{tcolorbox}


The resulting agent implementations are embedded into a LangGraph workflow with shared typed state, prompt
templates, and routing logic.
The population process combines mined process semantics with software architecture constraints: the mined artifacts determine what kinds of role behavior are meaningful. The LangGraph implementation turns those behaviors into executable agent nodes with explicit interfaces, routable responsibilities, and bounded actions.

\section{Evaluation}
\label{sec:evaluation}
We evaluated \approach through (1) a case study, (2)  functional testing of agent implementations, (3) a user study assessing the generated agent specifications.

\subsection{Case Study: Commitizen
}
\label{sec:cs}


We applied \approach to a repository of a real-world software project to demonstrate the technical feasibility of our approach.
We chose \textit{Commitizen}\footnote{\scriptsize \url{https://github.com/commitizen-tools/commitizen}, accessed 2026-06-05}, a well-established open-source project active since 2017 and having 3,400+ stars and 339 forks.

\smallskip
\noindent\emph{Step \ding{202} Event Log Extraction.}
The raw OCEL extracted for \textit{Commitizen} covers the period November 2017 -- November 2025
(approximately eight years of project history) and contains 21,488 events
and 4,813 objects across three object types: 1,459 issues, 2,765 commits, 
and 589 user accounts. Although the full log contains 21,488 events, the distribution is highly skewed. The \texttt{issue\_reporter} and \texttt{bot} roles account for most events. Consequently, generated specifications for low-volume roles should be interpreted as more exploratory.

\smallskip
\noindent\emph{Step \ding{203} Mapping to SE Tasks.}
Of the 2,765 commits in the log, {1,721 (62.2\%)} carry a well-formed
Conventional Commit message and were successfully enriched, yielding
1,721 task objects and a final OCEL of {6,534 objects} across four types
(commits, tasks, issues, users).

\smallskip
\noindent\emph{Step \ding{204} Determine Roles.}
To partition the enriched OCEL into role-specific sub-logs, \approach automatically assigns each of the
589 user objects to one of eight roles using a rule-based classifier applied in
priority order, see Table~\ref{tab:roles}.
The classifier first separates GitHub bot accounts (\texttt{bot}), then distinguishes
users who participate in issue discussions but never commit (\texttt{issue\_reporter})
from those who do.
Among committing users, \texttt{maintainer} is assigned to those with at least 20 commits
spanning at least three distinct task classes, reflecting broad, sustained involvement. The thresholds were selected as pragmatic heuristics for this proof-of-concept to distinguish sustained, diverse contribution from low-volume specialist activity. They were not optimized statistically and should be treated as configurable parameters rather than universal constants.
The remaining committing users are classified by their dominant task with \texttt{contributor} being a "catch-all" for low-volume or single-contribution users.
The four main, distinct roles (with $\geq$ 20 users) that remain are highlighted in bold.

\begin{table}[h]
\caption{Role distribution in the Commitizen event log.}

\label{tab:roles}
\centering
\scriptsize
\begin{tabular}{lrrr}
\hline
\textbf{Role} & \textbf{Users} & \textbf{Events} & \textbf{Objects}  \\
\hline
\textbf{issue\_reporter}    & 425 & 16,037 & 1,863  \\
\textit{contributor}        &  66 &    236 &   495  \\
\textbf{quality\_engineer}  &  37 &    194 &   395  \\
\textbf{technical\_writer}  &  23 &     71 &   194  \\
\textbf{feature\_developer} &  20 &    100 &   212  \\
maintainer         &   8 &  2,149 & 4,386  \\
\textit{bot}       &   6 &  3,394 & 1,055  \\
devops\_engineer   &   4 &     15 &    39  \\
\hline
\textbf{Total}     & 589 & 21,488 & 6,534  \\
\hline
\end{tabular}
\end{table}

\smallskip
\noindent\emph{Step \ding{205} Split Log.}
Given the above roles, the annotated OCEL was split accordingly, resulting in role-specific sub-logs, one per role.

\smallskip
\noindent\emph{Step \ding{206} Generate Agent Specification.}
Except \texttt{bot} and
\texttt{issue\_reporter}, all roles only performed a single activity type, and
thus DECLARE constraints are only informative for the above two roles. This does not invalidate the pipeline, but indicates that the utility of imperative and declarative mining depends on activity diversity in the role-specific sub-log.
The following is a summary of the markdown generated for the \texttt{issue\_reporter} role.

\begin{tcolorbox}[colback=gray!30, colframe=gray!90,
left=3pt, right=3pt, top=3pt, bottom=3pt]
\scriptsize
{\sffamily
\textbf{PROCESS DESCRIPTION:}
\texttt{ISSUE\_REPORTER}

\textbf{Overview:}
The \texttt{issue\_reporter} role initiates,
coordinates, and follows issue lifecycles,
linking issue and user objects through creation,
discussion, review, subscription, assignment,
closure, and reopening.

\textbf{Main Activities:}
Issue initiation (\texttt{created},
\texttt{labeled}, \texttt{assigned}),
communication (\texttt{commented},
\texttt{mentioned}, \texttt{subscribed}),
review handling (\texttt{review\_requested},
\texttt{reviewed},
\texttt{ready\_for\_review}),
lifecycle management (\texttt{closed},
\texttt{reopened}), and structural updates
(\texttt{referenced},
\texttt{cross\_referenced},
\texttt{renamed}).

\textbf{Object Interactions:}
The process centers on \texttt{issue} and
\texttt{user} objects. Issues are updated through
creation, discussion, and resolution, while users
participate as actors, reviewers, and collaborators.
Interactions coordinate visibility, accountability,
and traceability.

\textbf{Flow Patterns:}
Typical flows include
\texttt{created $\rightarrow$ commented},
repeated communication
(\texttt{commented $\rightarrow$ commented},
\texttt{commented $\rightarrow$ mentioned}),
review paths
(\texttt{reviewed $\rightarrow$ merged
$\rightarrow$ closed}),
and non-linear behaviors such as reopening or
review removal.

\textbf{Characteristics:}
Communication-driven, highly collaborative,
iterative, issue-centric, lifecycle-spanning,
and coordination-oriented.

\textbf{Business Interpretation:}
The role supports software delivery by ensuring
issues are captured, discussed, coordinated, and
resolved while maintaining visibility and
traceability.

}
\end{tcolorbox}

For the remaining roles, the OC-DFG is the primary
process model represented as shown in the following example, expressing that \texttt{Existence1[closed]} requires the activity \texttt{closed} occuring \textit{at least once}.

\begin{tcolorbox}[colback=gray!30, colframe=gray!90, left=3pt, right=3pt, top=3pt, bottom=3pt]
\scriptsize
\begin{verbatim}
{  "raw": "Existence1[closed] | |",
  "template": "Existence1",  "activity_a": "closed", "activity_b": ""  }
\end{verbatim}
\end{tcolorbox}


\smallskip
\noindent\emph{Step \ding{207} Generate LangGraph Agent Code.}
This yielded an issue-centric agentic application in which the \texttt{issue} object coordinates requests across all lifecycle stages. BPMN artifacts provided the issue management structure, DECLARE artifacts contributed policy checks and approval gating, and process descriptions populated the role-specific agents. 

The LangGraph application was generated by combining prompt interpretation with artifacts mined from the repository process rather than being handcrafted from scratch. A central design choice was that not every mined role became an agent.
Instead, the final implementation selected roles whose behavior was both sufficiently
distinct and operationally useful in a GitHub issue workflow.
This produced five role-specific agents: \texttt{issue\_reporter\_agent},
\texttt{bot\_workflow\_agent}, \texttt{implementation\_agent}, \texttt{quality\_agent},
and \texttt{technical\_writer\_agent}. The \texttt{implementation\_agent} combines the behavior mined from the roles: feature developer, contributor, maintainer, and DevOps engineer. Rather than exposing several narrowly separated development roles, the app consolidated them into a single agent 
responsible for interpreting feature requests, enhancements, and bug fixes as implementation work. The consolidation was deliberate: the mined roles showed similar implementation semantics, and the application required a practical executable role rather than a faithful organizational replica.

The example below shows a human-readable agent specification, also used in our user study.

\begin{tcolorbox}[colback=gray!30, colframe=gray!90,
left=3pt, right=3pt, top=3pt, bottom=3pt]
 \vspace{-.3em}
\scriptsize
{\sffamily
\textbf{SE AGENT: issue\_reporter}

\textbf{Goal: }You focus on intake, clarification, communication, and sustained issue coordination.

\textbf{Tasks (DO):}
• summarizing issue intent and missing context
• asking clarifying questions
• detecting duplicates
• suggesting labels, assignees, milestones
• preparing issues for downstream work
      
\textbf{Avoid (DON’T):}    
• make deep code changes directly
• close issues without strong resolution evidence

\textbf{Actions:}    
• “summarize\_issue”, “ask\_for\_clarification”
• Input: question: "Please provide expected behavior and actual behavior."
• Output: issue\_id, prepared documentation work items
}
\end{tcolorbox}

\subsection{Functional Testing}
\label{sec:test}
We tested the generated LangGraph application for \textit{Commitizen} using actual GitHub issues.
We conducted three smoke tests, selected to exercise distinct routing paths. These tests are not intended as evidence of general runtime effectiveness, but as sanity checks that the generated LangGraph application can ingest issues, route them, and produce coherent next actions.
In all cases, the app ingested the issue, loaded its GitHub context, interpreted the
text through LLM-based classification, routed the issue to a role-specific agent, and
derived a corresponding next action.

\textit{Issue~\#1: ``Readme.md requires more content to describe this project.''} 
This was interpreted as a documentation-oriented request.
The routing logic correctly selected the \texttt{technical\_writer\_agent}.
In turn this agent then used its LLM-based
analysis to derive a suitable documentation task focused on updating or preparing explanatory material.

\textit{Issue~\#2: ``Add a first basic hello-world function in Python with no input arguments
that prints `hello world'.''} 
This issue was routed correctly to the \texttt{implementation\_agent}.
The agent translated the request into a development-oriented action plan, containing a concrete implementation workflow comprising repository inspection, minimal code changes, traceability links to the originating issue, and preparation of a pull request for review. 

\textit{Issue~\#3: ``Application crashes when uploading large CSV files.''} This issue was routed correctly to the \texttt{bot\_agent}. The agent translated the issue into a workflow-oriented coordination plan, containing low-risk repository maintenance actions comprising issue triage through the application of \texttt{triage} and \texttt{needs-review} labels, routing to the appropriate review stage, and automatic reviewer assignment to the \texttt{maintainer} for further technical assessment.

\subsection{Exploratory User Study}
\label{sec:user_study}
To empirically evaluate \approach, we performed an exploratory, qualitative user study focusing on the four main, distinct roles identified in Table~\ref{tab:roles}.
The aim was to assess how well process-mined, project-specific agents address the problem of \textit{Human-AI alignment}, as perceived by the participants. \textit{Human-AI alignment} is a concept from HCI research referring to how well an AI agent meets the goal of producing the desired outcomes, without undesirable side effects~\cite{christian2020alignment,GoyalCT24}. 
Due to the project-specific separation of concerns and agent modularity, \approach's agent specifications should exhibit a relatively high degree of human-AI alignment.

We involved ten participants recruited from our pool of PhD students, teaching assistants, and Master students. Participants evaluated written specifications, not the running agents.


\vspace{-0.6em}
\subsubsection*{Study Flow:} The study entailed the following steps.
\textit{1. Motivation and Introduction:} Participants received a briefing on the problem of modeling and realizing \teammate{}s, the ensuing challenges of human-AI alignment, and the \approach approach.
\textit{2. Case Study \& User Study Goals:} 
    The participants were given an overview of the  case study, the generated SE Agent roles  and the evaluation dimensions (explained below).
    We shared all identified roles except \texttt{bot}, as this is already automated by Github.
\textit{3. The Survey:} 
    Human participants were informed about the study goals, protection of personal data and the voluntary nature of the study.
    Each participant received a set of agent specifications together with a questionnaire to be answered as detailed below.
    
\vspace{-0.6em}
\subsubsection*{Questionnaire Structure:}
We evaluate Human-AI alignment from a human-computer-interaction (HCI) perspective.
We address the following alignment concerns introduced in~\cite{GoyalCT24}, thereby contributing to the challenge of effective AI-human collaboration~\cite{RoychoudhuryPPR26}.
Assessments for these concerns are given on a five-point Likert scale: (1) strongly \underline{dis}agree, $\ldots$, (5) strongly agree.

\textit{Knowledge Schema:} The information needed by an agent for a successful transaction must be clear to prevent back-and-forth communication.
    Herewith, we assess what level of contextual clarity role-scoped agents may provide.

\begin{tcolorbox}[colback=gray!30, colframe=gray!90, left=3pt, right=3pt, top=2pt, bottom=2pt]
\scriptsize
{C1: The SE Agent specification clearly communicates what specific input the agent needs to successfully complete its task.}
\end{tcolorbox}

\textit{Autonomy:} Humans must understand the boundaries within which an AI agent can operate autonomously and when it must consult a human for decisions. 
\begin{tcolorbox}[colback=gray!30, colframe=gray!90, left=3pt, right=3pt, top=2pt, bottom=2pt]
\scriptsize
{C2: The SE Agent specification makes it clear when the agent is allowed to act autonomously and when it must wait for human approval or interaction.}
\end{tcolorbox}

\textit{Operational:} Humans and agents must actively identify and align on which tasks to execute, which requires that humans are aware of the agent skills.
\begin{tcolorbox}[colback=gray!30, colframe=gray!90, left=3pt, right=3pt, top=2pt, bottom=2pt]
\scriptsize
{C3: The tasks and responsibilities of the SE Agent are clearly identifiable, making it easy to understand what specific SE work I can assign to it.}
\end{tcolorbox}

\textit{Accountability:} Humans might worry that an agent misbehaves, which could negatively impact the humans' personal reputation.
    Consequently, for the sake of accountability, it must be possible to efficiently debug the AI agents.
    \begin{tcolorbox}[colback=gray!30, colframe=gray!90, left=3pt, right=3pt, top=2pt, bottom=2pt]
\scriptsize
{C4: If the SE Agent makes a mistake (e.g., merging bad code), the agent specification(s) allow me to isolate, understand, and debug the specific problem.}
\end{tcolorbox}
       
\textit{Human Engagement:} Because humans may have varying preferences for control and oversight, humans and SE agents should align on when, how, and why the human should be interrupted or engaged.

   \begin{tcolorbox}[colback=gray!30, colframe=gray!90, left=3pt, right=3pt, top=2pt, bottom=2pt]
\scriptsize
{C5: Coordinating hand-offs, monitoring outputs, and intervening with the SE Agent would require minimal cognitive effort and cause low friction.}
\end{tcolorbox}

To strengthen the findings and compensate for the relatively small sample size, we asked additional open-ended, cross-cutting questions.

\begin{tcolorbox}[colback=gray!30, colframe=gray!90, left=3pt, right=3pt, top=2pt, bottom=2pt]
\scriptsize

\textbf{Q1: \textit{Granularity}} -- Based on the agent specifications you reviewed, would you merge some agent roles, split them further, or keep them as generated?


\textbf{Q2: \textit{Collaboration}} -- Imagine you are a software engineer working alongside the AI agents on a real project. 
Are there any concrete moments in your workflow where you anticipate a handoff or coordination problem?
    If so, explain what you think which agent(s) and which reason(s) would cause it.


\textbf{Q3: \textit{Trust}} -- The role boundaries and SE agent behavior were  automatically derived rather than designed by a human. Are there potential reasons that would undermine your trust or acceptance of such generated SE agents?
\end{tcolorbox}

\vspace{-0.7em}
\subsubsection*{Data Quality and Threat Mitigation:}
To ensure the integrity of the survey data and mitigate the risk of careless responding (e.g., straight-lining), we implemented several  established quality control mechanisms: (1) To mitigate fatigue and ordering effects when answering four roles $\times$ eight questions (= 32 questions in total), the presentation order of the agent specifications was \textit{randomized} for each participant; (2) Statement C5 was \textit{negatively framed} to detect straight-lining; (3)    Responses that are \textit{logically inconsistent} (e.g., answering "Strongly Agree" to both C1 and the reverse-coded C5) were flagged.

    

\vspace{-0.5em}
\subsubsection{Results:} 
\begin{figure}[htbp]
     \centering
     \begin{subfigure}[b]{0.4\textwidth}
         \centering
         \includegraphics[width=\textwidth]{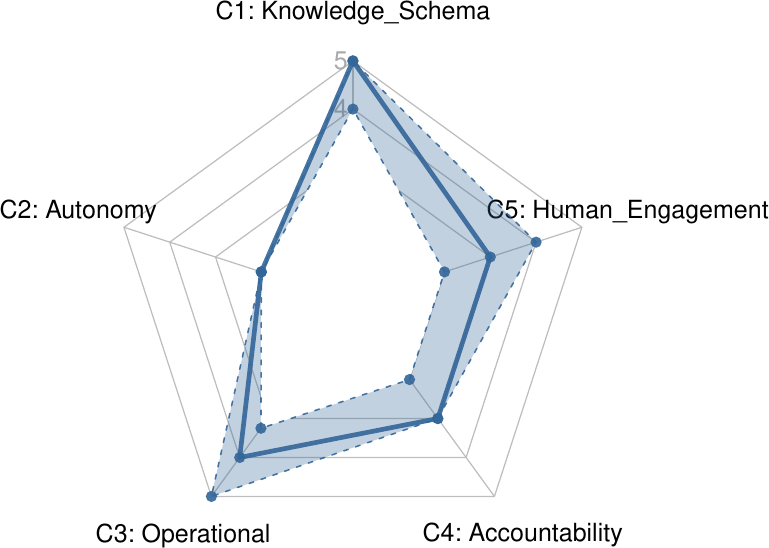}
         \caption{\texttt{quality\_engineer}} 
     \end{subfigure}
     \begin{subfigure}[b]{0.4\textwidth}
         \centering
         \includegraphics[width=\textwidth]{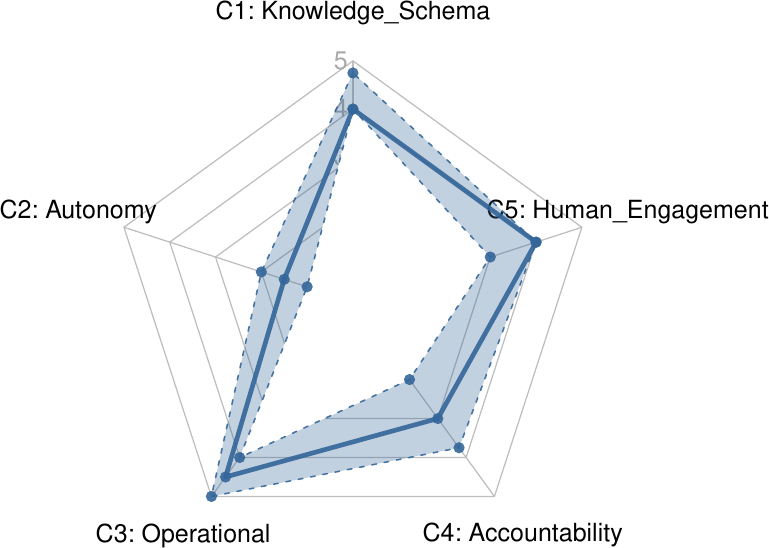}
         \caption{\texttt{technical\_writer}} 
     \end{subfigure}
\\
    \vspace{0.5em}
     
    
     \begin{subfigure}[b]{0.4\textwidth}
         \centering
         \includegraphics[width=\textwidth]{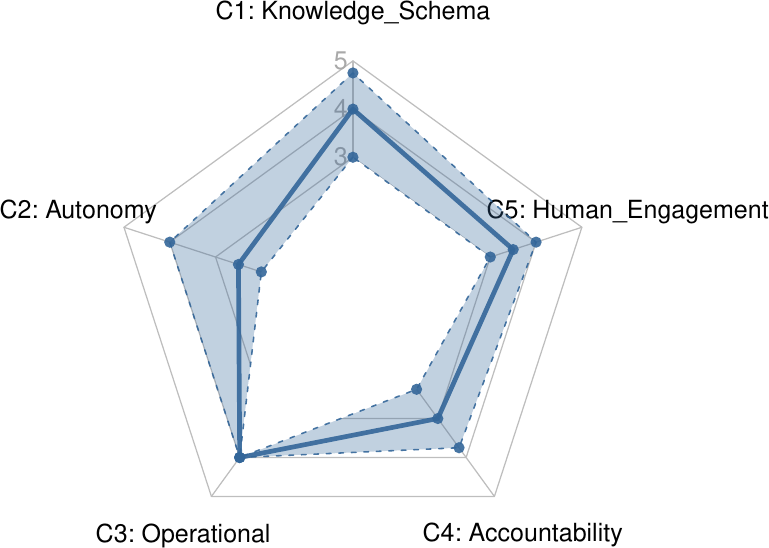}
         \caption{\texttt{issue\_reporter}} 
     \end{subfigure}
     \begin{subfigure}[b]{0.4\textwidth}
         \centering
         \includegraphics[width=\textwidth]{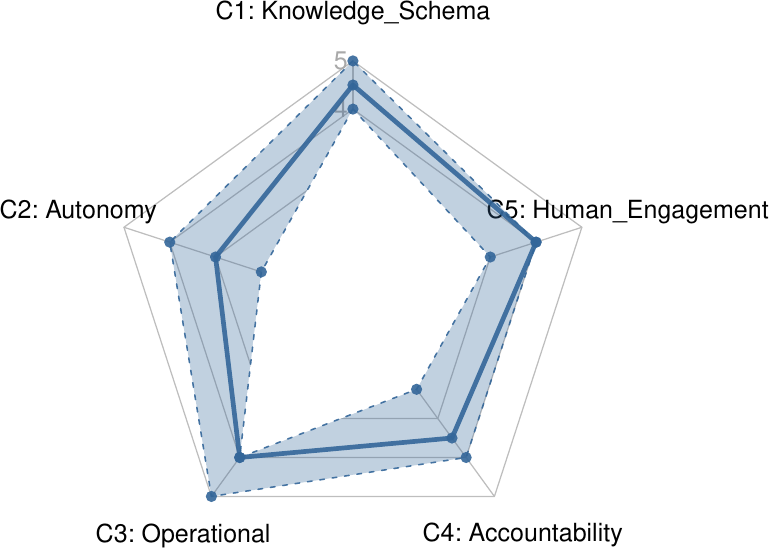}
         \caption{\texttt{feature\_developer}} 
     \end{subfigure}

 \caption{Radar Charts of the Qualitative User Study Results (showing Median as thick line and IQR as shaded area)}
     \label{fig:radar}
\vspace{-0.5em}
\end{figure}

Fig.~\ref{fig:radar} shows the medians of the participants' assessment for each of the four SE agents.
    
\vspace{-0.5em}
\subsubsection*{Overall Achievement of Alignment Criteria:}
Analyzing the medians across all four evaluated agent roles reveals distinct strengths and weaknesses in how process-mined specifications address human-AI alignment:

\textit{C1: Knowledge Schema (Med: 4):} This was the highest-rated dimension, indicating strong agreement that the specifications clearly communicate the specific inputs the agents need to successfully complete their tasks. The object-centric process models successfully delineate task scope and required entities.

\textit{C3: Operational Clarity (Med: 4):} Participants found the tasks and responsibilities of the agents to be highly identifiable, suggesting that deriving role boundaries from event logs yields practical and understandable work assignments.

\textit{C5: Human Engagement (Med: 4):} A moderate-to-high score implies that humans feel coordinating hand-offs, monitoring outputs, and intervening with the agents would require relatively low cognitive effort.

\textit{C4: Accountability (Med: 3):} The ability to isolate, understand, and debug an agent's mistakes scored neutrally. While process boundaries are clear, the underlying execution may still appear a ``black box'' for debugging.
    
\textit{C2: Autonomy (Med: 2):} This dimension scored notably low, highlighting a severe challenge. Humans struggled to understand the boundaries dictating when an agent is allowed to act autonomously versus when it must wait for human approval. This reflects a marked human-AI divergence on autonomy boundaries.

\vspace{-0.5em}
\subsubsection*{Role-Specific Assessment of C1-C5:}

A detailed breakdown of the five criteria across the four specific agent roles uncovers variations in how alignment is perceived depending on the nature of the software engineering tasks.

\texttt{feature\_developer:} This role exhibited the highest overall human alignment. Participants found its operational scope (4) and required knowledge schema (4.5) exceptionally clear. 
It also achieved the highest accountability score (3.5), likely because code implementations provide concrete, verifiable artifacts (e.g., a pull request) that are native to human review processes.

\texttt{quality\_engineer:} While the inputs (5) and tasks (4) for testing were well understood, accountability (3) and autonomy (2) dropped significantly, suggesting users are uncertain about how much leeway a testing agent has to automatically reject code or define its own test parameters.

\texttt{technical\_writer:} This role exposed the most extreme disparity in the study. 
While participants clearly understood its operational purpose (4.50) and the inputs it requires (4.00), it received an exceptionally poor autonomy score (1.5). 
Users seem highly uncertain about whether documentation agents publish changes autonomously or strictly draft content for human approval.

\texttt{issue\_reporter:} The issue reporter role yielded the most balanced, albeit slightly lower, assessment. 
Because this role centers on communication, coordination, and repeated review cycles, its autonomy boundaries (2.5) and human engagement expectations (3.5) may be slightly more apparent to human counterparts compared to the testing and writing roles.

\vspace{-0.5em}
\subsubsection*{Summary of Q1-Q3:}
Below, we summarize the main positive (+) and negative ($-$) comments to the open-text questions.

{\textit{Q1:}}
    (+) Roles map to distinct SE responsibilities, avoiding monolithic complexity and provide good coordination balance without excessive friction.
    ($-$) \texttt{feature\_developer} and \texttt{quality\_engineer} roles seem too broad. 
    Handoffs lack detailed boundaries, and there is perceived redundancy between \texttt{technical\_ writer} and \texttt{issue\_reporter}.

\textit{Q2:}
    (+) Agents successfully handle tedious tasks, such as technical documentation and systematic test coverage improvements.
    ($-$) Ambiguous validation criteria pose a risk of endless loops between \texttt{feature\_developer} and \texttt{quality\_engineer}. Furthermore, the rapid generation of agent output risks turning the human developer into a vetting bottleneck.
    
\textit{Q3:}
    (+) There is strong support for agents handling unappealing peripheral work and preparatory phases, allowing humans to focus on major architectural decisions.
    ($-$) Echoing Q2, risks include endless validation loops, premature actions (e.g., reporters interrupting drafts), and severe human review bottlenecks due to high-volume outputs lacking prioritization guardrails.

\subsection{Threats to Validity and Limitations}
\label{sec:validity}
Our user study measures \emph{perceived} human-AI alignment based on reading
agent specifications, not actual task performance or agent behavior.
A specification may appear well-scoped and comprehensible while producing
misaligned behavior at runtime.
We partly mitigate this through functional testing in Sec.~\ref{sec:test}.

The absolute assessments of our user study limit causal claims.
Yet, determining a fair baseline for a comparative assessment is rather challenging in our case.
Monolithic or fine-grained agent baselines are obviously unfair.
Using canonical SE roles -- as done by current frameworks~\cite{HeTL25} -- while presenting realistic and feasible baselines, would significantly overlap with the generated roles.
This makes it difficult to use them as hidden baselines, which are required to not compromise the study's validity due to psychological effects that skew perceived alignment.  

The participant sample comprised Master's students, teaching assistants, and PhD researchers from a single university, recruited as a convenience sample. 
This may introduce social desirability bias.
Also, as non-practicing software engineers, participants may not truly reflect the perspectives of developers deploying hybrid human-AI SE teams in production. 

\approach was evaluated on a single open-source project, \textit{Commitizen}, which follows the Conventional Commits specification and exhibits a relatively clean event log. Generalization to projects lacking structured commit conventions, or to proprietary or enterprise repositories, remains an open question.

A further threat stems from the single-role assignment in Step~\ding{204}. Open-source contributors routinely act in multiple capacities, and forcing one label per user withholds behavior from the specialist sub-logs it could inform: in our case study, the maintainer role (8 users, 2,149 events) absorbs documentation, testing, and development activity that would otherwise enrich the \texttt{technical\_writer}, \texttt{quality\_engineer}, and \texttt{feature\_developer} models. Extending role assignment to overlapping memberships is future work.


SE event logs capture coarse-grained developer activities (e.g., commits or issue updates), whereas AI coding agents operate at a finer granularity (e.g.,  writing code or running tests). 
Mined role specifications thus may not fully capture deployed agent behavior. 
OCPM partly mitigates this by linking commits, issues, and pull requests across object types.

\section{Conclusion}
\label{sec:conclusion}




We presented \approach, a pipeline that derives project-specific SE agent specifications and implementations from SE repository event logs by mining roles, behavioral patterns, and normative constraints. 
As proof-of-concept we realized the pipeline and applied it to the \textit{Commitizen} open-source project.
Initial smoke tests showed coherent issue routing, and an exploratory user study suggested that the generated specifications appear to be well-aligned with human expectations.


Future work will evaluate \approach on more diverse repositories, including projects without structured commits and proprietary code bases, involve practicing engineers, and extend role assignment from hard partitions to overlapping role memberships.
Enriching event logs with IDE telemetry and CI/CD logs will support finer-grained agent scoping and specifications.


\bibliographystyle{splncs04}
\bibliography{tosem-refs,refs,infosys-refs,refs-tam}

\end{document}